# Electrical anisotropy of heteroepitaxial InSb/GaAs layers

T.A. Komissarova[1]*, A.N. Semenov[1], D.A. Kirilenko[1], B.Ya. Meltser[1], V.A. Solov'ev[1], A.A. Sitnikova[1], P. Paturi[2], and S.V. Ivanov[1]

[1]Ioffe Physical Technical Institute of Russian Academy of Sciences, St Petersburg 194021, Russia

[2]Wihuri Physical Laboratory, Department of Physics and Astronomy, University of Turku, FIN-20014, Turku, Finland

We report on study of electrical and structural properties of InSb/GaAs(001) heteroepitaxial layers in [110] and [1-10] crystallographic directions. Strong anisotropy of electron transport parameters measured at a low magnetic field has been found, whereas the electrical transport through the InSb bulk layer studied by Shubnikov-de Haas oscillations is shown to be independent of the crystallographic directions. The low-field electrical anisotropy of the InSb films appears to be governed by two competitive anisotropic effects: influence of spontaneously formed In nanoclusters inhomogeneously distributed within the InSb layers and conductivity through the near-interface layer with high anisotropic density of extended defects.

*) Author to whom correspondence should be addressed

electronic mail: komissarova@beam.ioffe.ru

InSb, being the narrowest bandgap III-V material, is widely used in such applications as infrared detectors and magnetic sensors, and is considered as a prospective candidate for power-consumed high-speed devices [1]. For the electronic devices, InSb epitaxial films are used to be grown on strongly lattice-mismatched semi-insulating GaAs(001) substrates and contain high density of extended defects (well above $10^7$ cm$^{-2}$) (see e.g. [2] and refs. therein). Knowledge and control of transport parameters of electrons in the InSb/GaAs epilayers are of great importance in terms of device fabrication. Electrical properties of InSb films have been widely studied for many decades, in particular, the geometrical magnetoresistance effect [3,4], influence of interface layer on electrical properties of InSb films [5-7], and inhomogeneity of electron mobility along growth direction [3,8] were observed.

Recently, anisotropy of electron mobility in InSb/Al$_x$In$_{1-x}$Sb quantum well (QW) structures grown on GaAs(001) has been found [9]. This was suggested to be due to electron scattering on micro-twin (MT) defects having anisotropic density in the [110] and [1-10] directions [9,10]. Then one can expect analogous anisotropy of the electron mobility in the bulk In(Al)Sb epitaxial layers grown on GaAs, since the same structural defects should exist there. Nevertheless, most of the published data on electrical properties of InSb/GaAs epilayers were obtained by using a Van der Pauw geometry, which is insensitive to crystallographic directions, while no data on lateral electrical anisotropy of InSb/GaAs(001) layers have been reported so far. But obviously this problem can be essential for electronic applications of the InSb epilayers.

In this Letter we report on electrophysical properties of InSb/GaAs(001) epitaxial layers, studied along the [110] and [1-10] crystallographic directions by resistivity and Hall effect measurements in the wide temperature (1.6-300)K and magnetic field (0-30T) ranges, in correlation with their structural properties revealed by using transmission electron microscopy (TEM). Anisotropy of Hall concentration and mobility of the InSb films measured at a low magnetic field has been observed. However, it has been found that the transport parameters of quantized electrons in the bulk of the InSb films are isotropic. Anisotropy of low-field parameters is shown to originate from two additional parasitic conductivity channels induced by spontaneously formed In nanoparticles and conductivity through a strongly defective InSb/GaAs near-interface layer, which depend on crystallographic directions.

We have studied InSb films grown by molecular beam epitaxy on GaAs (001) substrates on top of AlSb buffer layers. Thickness of the InSb epitaxial layers was in the range (1-3) μm. The AlSb buffer layers were grown at high temperatures (540-560ºC) to minimize the stacking faults (SFs) and MT density, as their formation has been established to be enhanced at low growth temperatures [11]. Thickness of the AlSb buffer layers was varied from 0.36 to 1 μm. Values of the electron concentration and mobility of the studied InSb layers, measured in Van der Pauw (VdP) geometry at magnetic field B = 0.2T at 300 K, are presented in Table I (column 5). All the other electrical measurements were performed in a Hall bar geometry with current flowing in [110] and [1-10] crystallographic directions. Measurements of the Hall coefficient $R_H$ and resistivity $\rho$ were carried out in the temperature range (1.6-300) K in pulsed magnetic fields up to 30 T. Structural properties of the InSb films were studied by TEM using Philips EM-420 ( accelerating voltage 100 kV) and Jeol JEM-2100F (acc. volt. 200 kV) equipped by energy-dispersive X-ray spectrometer (EDS). The cross-sectional TEM specimens were prepared in two mutually perpendicular orientations in order to reveal both [110] and [1-10] crystallographic directions..

Strong difference in the values of the Hall concentration and mobility measured at $B = 0.2$ T in the [110] and [1-10] crystallographic directions has been found in the studied InSb films (Table I, columns 6, 7). However, TEM measurements exhibit occurrence of SFs only in the bulk of InSb layers grown atop of the AlSb buffer with thickness below 500 nm (Fig.1). It allows us to assume that the observed electrical anisotropy does not relate to the anisotropic density of the SFs (or MTs) only. To elucidate origin of the strong electrical anisotropy in the InSb layers high-magnetic field measurements were performed.

Typical magnetic field dependences of the absolute values of the Hall coefficient $|R_H|$ and $\rho$ measured in the two crystallographic directions are presented in Fig.2. One can see that $|R_H|$ and $\rho$ versus B dependences differ in [110] and [1-10] directions. Magnetic field dependence of $|R_H|$ consists of three parts. First, $|R_H|$ increases with increasing the magnetic field, which is accompanied by practically linear increase of $\rho$. Simultaneously, Shubnikov-de Haas oscillations are observed at magnetic fields up to 5T at low temperatures ($T < 50$ K). Then $|R_H|$ starts to decrease and $\rho$ continues to increase with magnetic field by exponential (at $T = 4.2$ K) or linear (at higher temperatures) law.

The SdH oscillations extracted from the linear magnetic-field dependence of the resistivity exhibit a period which does not depend on the crystallographic directions (Fig. 3), i.e. the concentration of quantized electrons $n_{SdH}$ in the InSb layers is isotropic. $n_{SdH}$ values are smaller than the Hall concentration measured at $B = 0.2$ T in both [110] and [1-10] crystallographic directions ($n_I$ and $n_{II}$) (Table I, columns 6, 7, 8). It means that in addition to the bulk conductivity there exist other anisotropic conductivity mechanisms in the InSb layers which are responsible for the $|R_H|$ and $\rho$ magnetic field dependences and, hence, for the difference between $n_I$, $\mu_I$ and $n_{II}$, $\mu_{II}$.

Increasing magnetic-field dependence of $|R_H|$ cannot be explained in the frames of the conventional semiconductor theory, since it should decrease with magnetic field or remain constant in this case [12]. The absolute value of $R_H$ can increase with increasing magnetic field due to decreasing electron concentration in the conduction band in case of magnetically-induced freeze-out of donor impurity, which was studied in bulk InSb [13,14]. However, this quantum effect can be observed only at low temperatures ($< 10$ K) while the increasing $|R_H|$ vs $B$ dependences in the studied InSb epilayers are measured in the whole temperature range. Moreover, in the case of magnetically-induced freeze-out of donor impurity the Hall coefficient increases exponentially with magnetic field [14], which is not the case here. Therefore, the only possible explanation of the increasing magnetic-field dependences of $|R_H|$ in the InSb layers appears to be the influence of highly conducting inhomogeneities [15]. In this case, at low magnetic fields electrical current flows mainly through the conducting inclusions leading to measurements of understated values of the Hall coefficient and resistivity. Increasing of the magnetic field leads to pushing out the current flow lines from the highly conducting inhomogeneities and increase of $|R_H|$ and $\rho$.

Measurements of temperature dependences of the resistivity at T< 4 K as well as TEM-EDS analysis of specific structural defects in the InSb layers showed that metallic In nanoclusters is the most feasible origin of such inhomogeneities. In the temperature range of (3.5 - 3.7) K, the temperature dependences of $\rho$ reveal a kink, starting to decrease rapidly with temperature reduction. This effect corresponds to the onset of superconducting transitions in the metallic In nanoclusters. In accordance with the McMillan law, the temperatures of these transitions are higher than the critical temperature of bulk indium ($T_c$ = 3.41 K) due to small sizes of the superconducting In clusters [16]. Simultaneous analysis performed by TEM-EDS technique has demonstrated that some extended defects in the bulk of the InSb layers contain excess of metallic indium. It has been found that the density of such defects differs for [110] and [1-10] crystallographic directions. Probably, spontaneous formation of the metallic In nanoclusters in the InSb epilayers is caused by small InSb formation enthalpy (-30.1 kJ/mol) [17], low In-Sb bond energy (1.57 eV) as compared with Sb-Sb bonds (3.13 eV) [18], and high density of the extended defects which can accumulate excessive indium.

A decrease of the absolute value of the Hall coefficient with magnetic field along with positive magnetoresistance effect in a semiconductor film is usually observed in case of presence of a second conductivity channel [12]. It is reasonable to assume that the strongly disturbed InSb layer in the vicinity of the InSb/AlSb heterointerface ($\Delta a/a$~5.6%) can serve as the additional conductivity channel in the InSb films [5-7]. Indeed, plan-view TEM studies confirmed that the density of the extended defects (dislocations and SFs) in the near-interface region is much higher (>$10^{10}$ cm$^{-2}$) than that in the bulk of the InSb layers (Fig.4a,b). Moreover, plan-view TEM measurements of the region performed in (220) and (2-20) reflections reveal strongly different extended defect densities in the [1-10] and [110] crystallographic directions (Fig.4c,d), in particular, the SFs density varies from $4\times10^7$ to $2\times10^8$ cm$^{-2}$, respectively. This can lead to anisotropic conductivity of the layer and, hence, to its anisotropic contribution to the measured integral Hall coefficient and resistivity. It should be noted that electrons of the near-interface layer do not contribute to the SdH oscillations due to their fast scattering at the extended defects.

Exponential increasing of the resistivity observed only at low temperatures (T < 10 K) most likely corresponds to the extreme quantum limit [19,20], which should not affect the Hall effect measurements [21]. Additionally, at high magnetic fields we cannot exclude the geometrical magnetoresistance effect due to the contact size and arrangement [13].

Therefore, the values of the Hall concentration and mobility measured at a low magnetic field value in the [110] and [1-10] crystallographic directions are governed by at least two parasitic anisotropic conductivity effects: influence of the In nanoclusters and conductivity through the near-interface layer, rather than by the conductivity through the bulk of the InSb films. The similar experimental results were obtained for Al$_{0.09}$In$_{0.91}$Sb epitaxial layers, which allows one to assume that the electrical anisotropy observed in the InSb/AlSb/AlInSb heterostructures [9,10] could be related to the anisotropy of the AlInSb barrier layers rather than the 2DEG channel.

In conclusion, strong anisotropy of the electron concentration and mobility of the InSb/GaAs(001) heteroepitaxial layers, measured at a low magnetic field value in the [110] and [1-10] crystallographic directions, has been found. Electrical transport through the bulk of the InSb films is shown to be isotropic. In contrast, there exist two competitive anisotropic effects: influence of the inhomogeneous distribution of In nanoparticles and conductivity through the near-interface layer with high laterally-anisotropic density of the extended defects, which are responsible for the difference in the low-field transport parameters of electrons in the [110] and [1-10] directions.

Acknowledgements. The work was partly supported by RFBR Grant #11-02-12249-ofi-m. Jenny and Antti Wihuri Foundation is acknowledged for financial support.

Table I. Electrical and structural parameters of InSb layers

| Sample | d, μm | $N_{disl}$, cm$^{-2}$ | T, K | VdP $n_H$, cm$^{-3}$ | VdP $\mu_H$, cm$^2$/Vs | [1-10] direction $n_I$, cm$^{-3}$ | [1-10] direction $\mu_I$, cm$^2$/Vs | [110] direction $n_{II}$, cm$^{-3}$ | [110] direction $\mu_{II}$, cm$^2$/Vs | $n_{SdH}$, cm$^{-3}$ |
|---|---|---|---|---|---|---|---|---|---|---|
| 1 | 2 | 3 | 4 | 5 | | 6 | | 7 | | 8 |
| A | 1.3 | (1–1.5)×10$^9$ | 300 | 2.5×10$^{16}$ | 48500 | 7×10$^{16}$ | 18100 | 6.6×10$^{17}$ | 2650 | |
| A | 1.3 | (1–1.5)×10$^9$ | 4.2 | | | 4.3×10$^{16}$ | 17600 | 1.3×10$^{17}$ | 2800 | 4.1×10$^{16}$ |
| B | 1.8 | 3.5×10$^8$ | 300 | 4.5×10$^{16}$ | 32200 | 4.6×10$^{16}$ | 26200 | 1.8×10$^{17}$ | 6100 | |
| B | 1.8 | 3.5×10$^8$ | 4.2 | | | 2.6×10$^{16}$ | 27700 | 1.2×10$^{17}$ | 5700 | 2.5×10$^{16}$ |
| C | 2.25 | (2–3)×10$^8$ | 300 | 3.6×10$^{16}$ | 34800 | 3.1×10$^{16}$ | 55000 | 6.5×10$^{16}$ | 16700 | |
| C | 2.25 | (2–3)×10$^8$ | 4.2 | | | 4×10$^{16}$ | | 2.9×10$^{16}$ | 16000 | 1.4×10$^{16}$ |

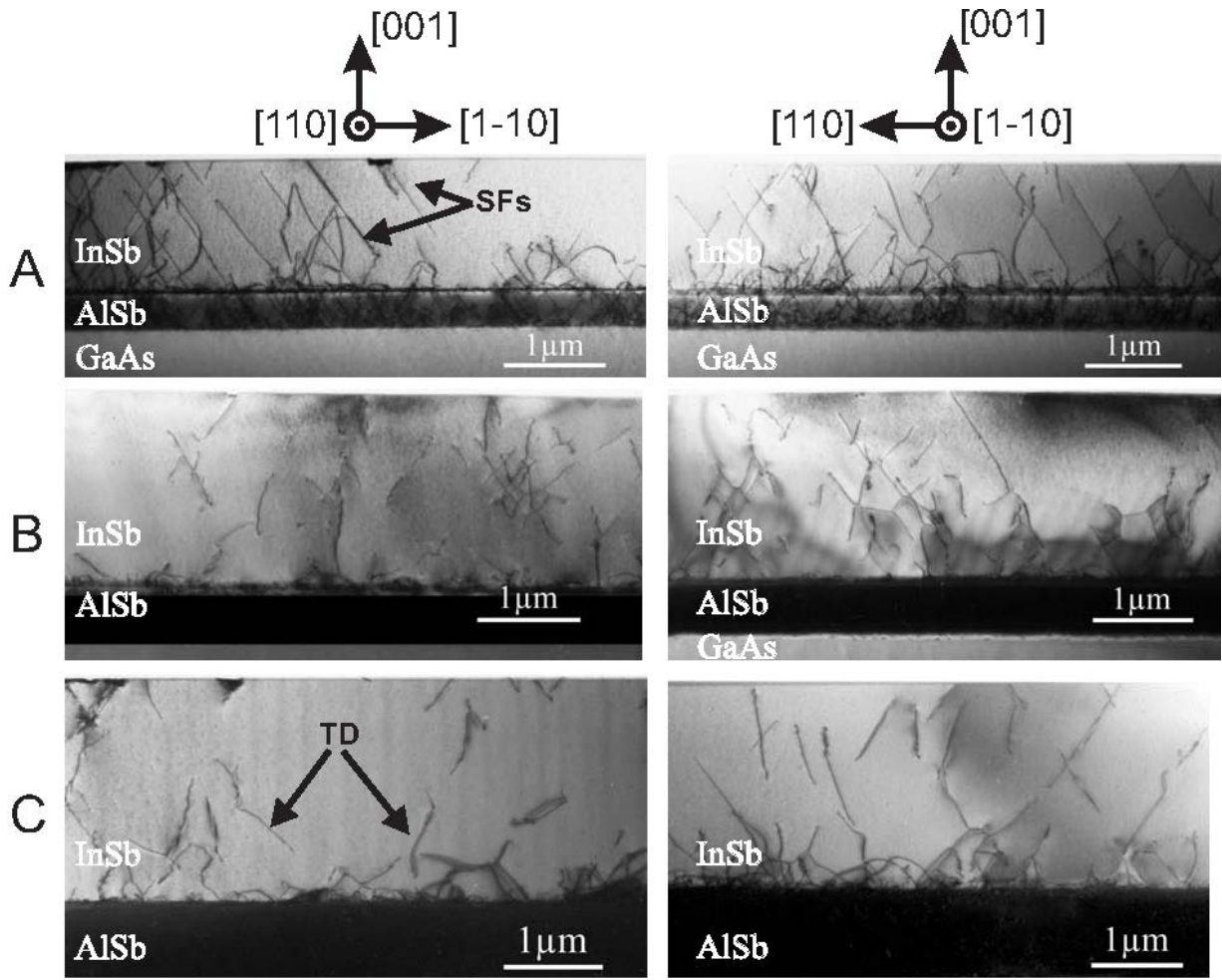

Fig.1. Cross-sectional TEM images of three InSb/GaAs(001) layers grown atop of different AlSb buffer layers (correspond to Table I): A – $d_{AlSb}$ = 360 nm, $T_S$ = 560ºC; B – $d_{AlSb}$ = 500 nm, $T_S$ = 540ºC; C – $d_{AlSb}$ = 1000 nm, $T_S$ = 560ºC. For each layer the left and right images were obtained on samples cleaved in two perpendicular directions [1-10] and [110], respectively. TD means threading dislocations.

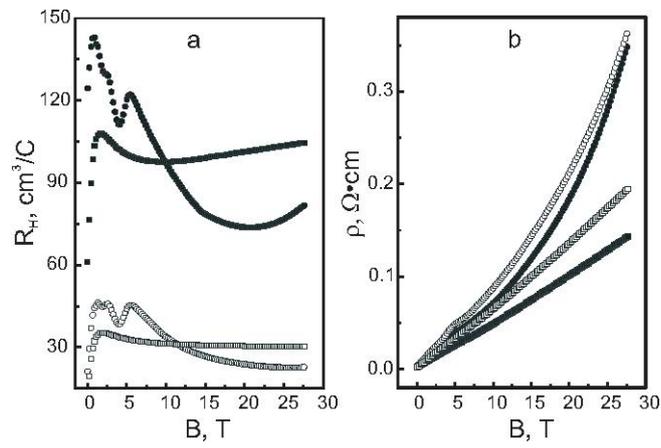

Fig. 2. Typical dependences of the absolute value of Hall coefficient (a) and resistivity (b) of the InSb layers measured at $T$ = 300 K (squares) and $T$ = 4.2 K (circles) in two crystallographic directions [1-10] and [110] (full and empty symbols, respectively).

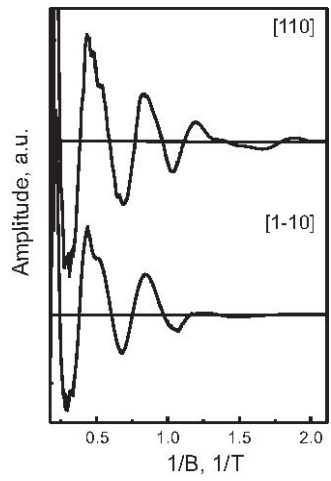

Fig. 3. Shubnikov-de Haas oscillations measured in one of the InSb layers in two crystallographic directions at T = 4.2 K.

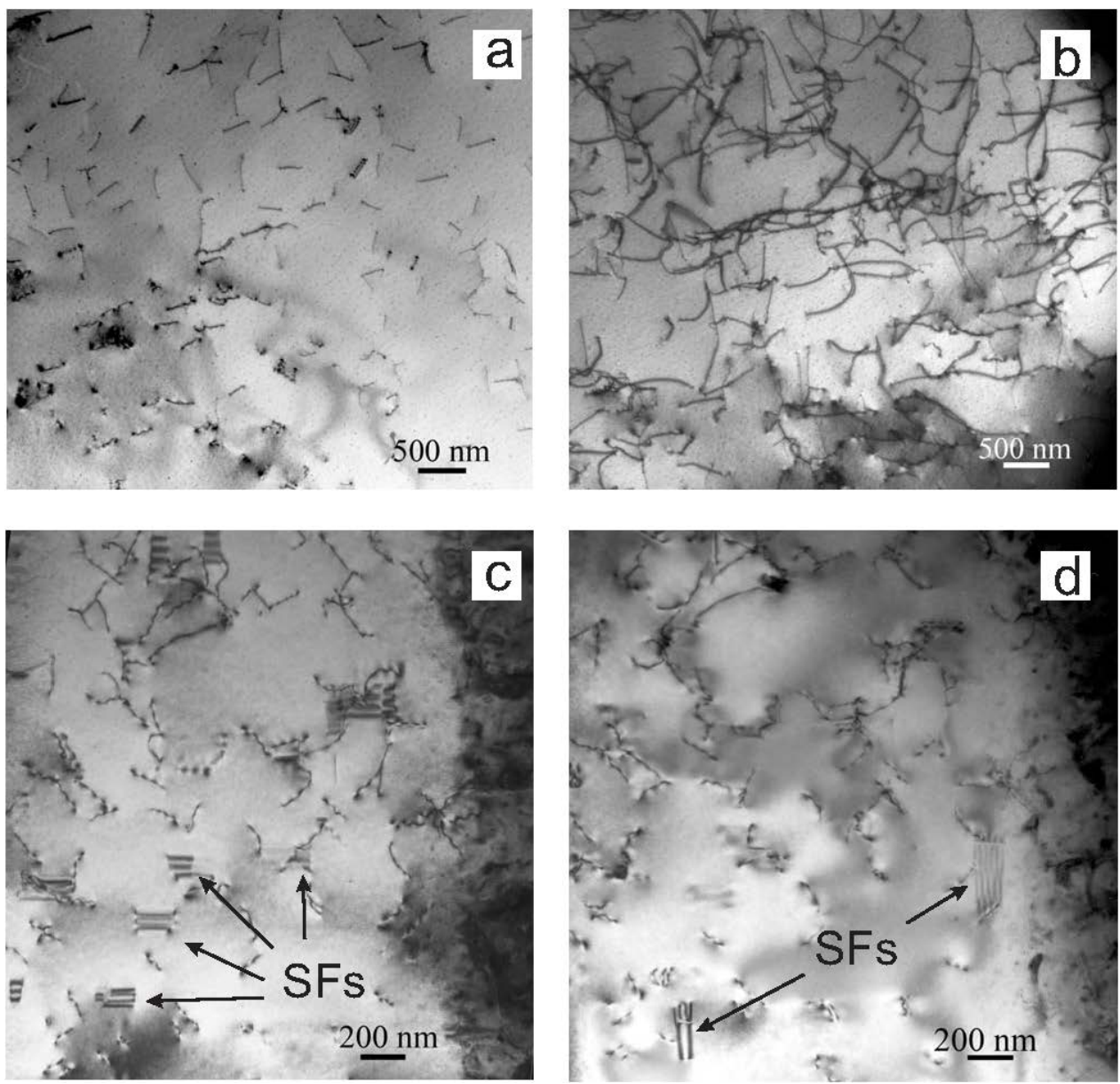

Fig. 4. Plan-view TEM images of the bulk (a) and the near-interface region (b) of sample A, as well as the images of the near-interface area of sample B, obtained in (220) (c) and (2-20) (d) reflections.